\documentclass[11pt,letterpaper]{article}%
\usepackage{amsmath}
\usepackage{epsfig}
\usepackage{graphicx}%
\usepackage{amsfonts}%
\usepackage{amssymb}
\begin{document}
%
%TCIMACRO{\TeXButton{titlepage}{\begin{titlepage}
%\begin{flushright}
%UCL-IPT-03-01
%\end{flushright}
%\vspace*{30mm}
%\begin{center}
%\huge{Are  $B\rightarrow\pi K$  CP-asymmetries quantized ?}
%\end{center}
%\vspace*{10mm}
%\begin{center}
%\Large{Jean-Marc G\'{e}rard\footnote{gerard@fyma.ucl.ac.be}
%and Christopher Smith\footnote{smith@fyma.ucl.ac.be}}
%\end{center}
%\vspace*{5mm}
%\begin{center}
%Institut de Physique Th\'{e}orique, Universit\'{e}
%catholique de Louvain\\
%Chemin du Cyclotron, 2, B-1348, Louvain-la-Neuve, Belgium
%\end{center}
%\vspace*{5mm}
%\begin{center}
%January 21, 2003
%\end{center}
%\vspace*{10mm}
%\begin{abstract}
%Search for patterns in the numerous B-decay modes now available is
%necessary in order to test the Cabibbo-Kobayashi-Maskawa theory of
%CP-violation. In particular, the well-structured pattern of $B\rightarrow\pi
%K$ branching
%ratios may lead to a quantized spectrum for direct CP-asymmetries, providing
%in this way a rather unique
%opportunity to discriminate between hadronic final state interaction models.
%\end{abstract}
%\end{titlepage}}}%
%BeginExpansion
\begin{titlepage}
\begin{flushright}
UCL-IPT-03-01
\end{flushright}
\vspace*{30mm}
\begin{center}
\huge{Are  $B\rightarrow\pi K$  CP-asymmetries quantized ?}
\end{center}
\vspace*{10mm}
\begin{center}
\Large{Jean-Marc G\'{e}rard\footnote{gerard@fyma.ucl.ac.be}
and Christopher Smith\footnote{smith@fyma.ucl.ac.be}}
\end{center}
\vspace*{5mm}
\begin{center}
Institut de Physique Th\'{e}orique, Universit\'{e}
catholique de Louvain\\
Chemin du Cyclotron, 2, B-1348, Louvain-la-Neuve, Belgium
\end{center}
\vspace*{5mm}
\begin{center}
January 21, 2003
\end{center}
\vspace*{10mm}
\begin{abstract}
Search for patterns in the numerous B-decay modes now available is
necessary in order to test the Cabibbo-Kobayashi-Maskawa theory of
CP-violation. In particular, the well-structured pattern of $B\rightarrow\pi
K$ branching
ratios may lead to a quantized spectrum for direct CP-asymmetries, providing
in this way a rather unique
opportunity to discriminate between hadronic final state interaction models.
\end{abstract}
\end{titlepage}%
%EndExpansion

\newpage

\section{Introduction}

In 2001, soon after the starting of two dedicated B-factories, CP-violation
was observed for the first time in B-meson decays, more precisely in the
interference of mixing and decay of $B^{0}-\bar{B}^{0}$ states, by the BaBar
Collaboration at SLAC \cite{BaBarCP} and the Belle Collaboration at KEK
\cite{BelleCP}. This rather spectacular result definitely promoted the
Cabibbo-Kobayashi-Maskawa (CKM) parametrization \cite{KM} to the first theory
of microscopic irreversibility.

In 1999, direct CP-violation was eventually established in K-meson decays by
two other experiments (NA48 at CERN \cite{NA48} and KTeV at Fermilab
\cite{KTeV}), after tremendous efforts. Today, the $\varepsilon^{\prime
}/\varepsilon$ parameter is accurately known experimentally, but is still of
no use to constrain the unitary CKM mixing matrix.

Strong dynamics at low scale is mainly responsible for this paradoxical
situation in weak decay physics. In particular, the so-called $\Delta I=1/2$
rule leads to large hadronic uncertainties in $K\rightarrow\pi\pi$ decay
amplitudes. At the B-mass scale, we expect genuine patterns to emerge for
branching ratios of two-body weak decay processes related by hadronic flavor
symmetries such as isospin.

So, the future of $B$-meson phenomenology may look bright in view of the
opening of so many decay channels. However, if we want to shed some new light
on CP-violation, we have somehow to rely on a quark-gluon picture. But the
Operator Product Expansion methodology inevitably involves many hadronic
matrix elements. Moreover, present technologies within QCD do not provide
rigorous predictions on the CP-conserving strong phases induced by final state
interactions (FSI) at the finite B-mass scale. Consequently, we usually have
to rely on specific models for hadron dynamics to probe the CKM mixing matrix
through our beloved direct CP-asymmetries.

Many bounds on the CKM angle $\gamma$ from $B\rightarrow\pi K$ decay rates and
CP-asymmetries have been derived since five years \cite{OtherWorks}. However,
they all require some assumption on the origin and size of the CP-conserving
strong phases involved. In this letter, we would like to illustrate, without
theoretical prejudice beyond isospin invariance, how the observed
$B\rightarrow\pi K$ decay modes might become soon a reliable test for the
rather simple though controversial SU(2)-elasticity and on-shell
$c\overline{c}$ rescattering hypotheses.

\section{Quantized CP-asymmetries in $K\rightarrow\pi\pi$}

The measured exclusive semi-leptonic decay width ratio \cite{PDG2002}%
\begin{equation}
\frac{\Gamma\left(  K^{+}\rightarrow e^{+}\pi^{0}\nu_{e}\right)  }%
{\Gamma\left(  K^{0}\rightarrow e^{+}\pi^{-}\nu_{e}\right)  }\simeq0.52
\label{Eq1}%
\end{equation}
is very close to the value expected from isospin symmetry. A factor 1/2
originates indeed from the $1/\sqrt{2}$ Clebsch-Gordan coefficient for
$\pi^{0}$. So, the large $K^{+}$ lifetime observed \cite{PDG2002}%
\begin{equation}
\frac{\tau\left(  K^{+}\right)  }{\tau\left(  K^{0}\right)  }\simeq70
\label{Eq2}%
\end{equation}
requires a dominance of the $\Delta I=1/2$ amplitude $A_{0}$ over the $\Delta
I=3/2$ amplitude $A_{2}$:%
\begin{equation}
\left|  \frac{A_{2}}{A_{0}}\right|  ^{\exp}\approx\frac{1}{22} \label{Eq3}%
\end{equation}
if we adopt the standard parametrization for the hadronic K-decay amplitudes%
\begin{align}
A\left(  K^{+}\rightarrow\pi^{0}\pi^{+}\right)   &  =\sqrt{\frac{3}{2}}%
A_{2}e^{i\delta_{2}}\nonumber\\
A\left(  K^{0}\rightarrow\pi^{-}\pi^{+}\right)   &  =\sqrt{\frac{2}{3}}%
A_{0}e^{i\delta_{0}}+\sqrt{\frac{1}{3}}A_{2}e^{i\delta_{2}}\label{Eq4}\\
A\left(  K^{0}\rightarrow\pi^{0}\pi^{0}\right)   &  =\sqrt{\frac{1}{3}}%
A_{0}e^{i\delta_{0}}-\sqrt{\frac{2}{3}}A_{2}e^{i\delta_{2}}\;.\nonumber
\end{align}

The CKM theory for CP-violation requires a link between the hadron world and
its quark-gluon representation. In the free-quark approximation, the effective
$\Delta S=1$ weak Hamiltonian for a semi-leptonic or hadronic decay simply
factorizes into two currents. However, hadronic decays also involve
non-factorizable gluon exchanges between these currents.

In particular, factorizable (F) quark diagrams easily explain Eq.(\ref{Eq1})
but are unable, alone, to produce the $\pi^{0}\pi^{0}$ final state (see
Fig.1a). Hence%
\begin{equation}
\left(  \frac{A_{2}}{A_{0}}\right)  ^{F}=\frac{1}{\sqrt{2}} \label{Eq5}%
\end{equation}
and this neutral state is only reachable through either $\pi^{-}\pi^{+}$
hadronic rescattering ($\delta_{0}\neq\delta_{2}$) or non-factorizable (NF)
quark diagrams (see Fig.1b).%

\begin{align*}
&
\begin{array}
[c]{cc}%
%TCIMACRO{\FRAME{itbpF}{2.7622in}{0.8077in}{0in}{}{}{fig1a.eps}%
%width 2.7622in;  height 0.8077in;  depth 0in;  original-width 4.9718in;
%original-height 1.4183in;  cropleft "0";  croptop "1";  cropright "1";
%cropbottom "0";  filename 'Fig1a.eps';file-properties "XNPEU";}}}%
%BeginExpansion
{\includegraphics[
height=0.8077in,
width=2.7622in
]%
{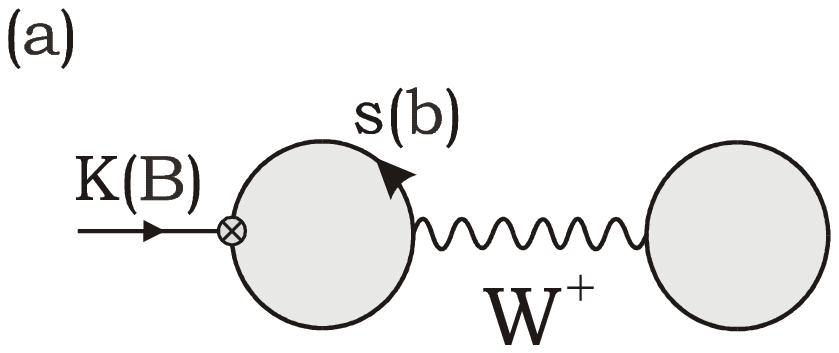}%
}%
%EndExpansion
&
%TCIMACRO{\FRAME{itbpF}{1.6924in}{0.9478in}{0.1487in}{}{}{fig1b.eps}%
%width 1.6924in;  height 0.9478in;  depth 0.1487in;  original-width 3.0277in;
%original-height 1.6743in;  cropleft "0";  croptop "1";  cropright "1";
%cropbottom "0";  filename 'Fig1b.eps';file-properties "XNPEU";}}}%
%BeginExpansion
\raisebox{-0.1487in}{\includegraphics[
height=0.9478in,
width=1.6924in
]%
{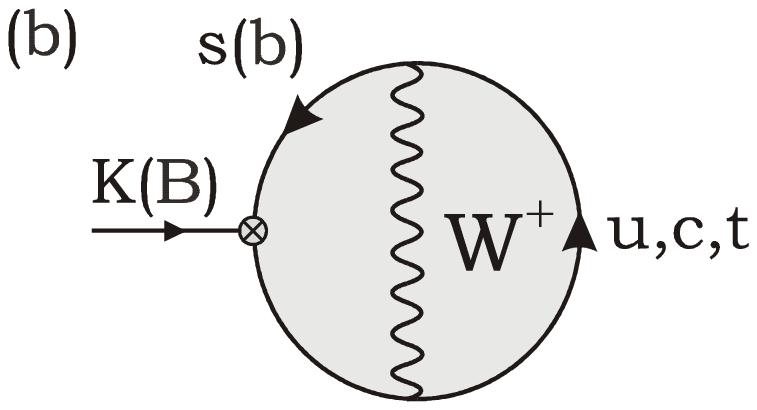}%
}%
%EndExpansion
\end{array}
\\
&  \underline{Fig.1}\text{\ To first order in }G_{F}\text{ and to all orders
in }\alpha_{S}\text{: }(\text{a})\text{ factorizable and }(\text{b})\text{
non-}\\
&  \;\;\;\;\;\;\;\text{factorizable }\Delta S\left(  \Delta B\right)
\overset{}{=}1\text{ quark diagrams, before final state hadronization.}%
\end{align*}

Nowadays, we are convinced that strong dynamics is indeed fully responsible
for the empirical $\Delta I=1/2$ rule \cite{PDG2002}%
\begin{align}
\frac{Br\left(  K_{S}\rightarrow\pi^{0}\pi^{0}\right)  }{Br\left(
K_{S}\rightarrow\pi^{-}\pi^{+}\right)  }  &  =0.458\pm0.004\nonumber\\
&  \approx\frac{1}{2}\left\{  1-3\sqrt{2}\operatorname{Re}\left(  \frac{A_{2}%
}{A_{0}}\right)  \cos\left(  \delta_{2}-\delta_{0}\right)  \right\}
\label{Eq6}%
\end{align}
and provides the necessary ingredients to generate non-zero $K\rightarrow
\pi\pi$ CP-asymmetries%
\begin{equation}
A_{CP}^{K}\left(  \pi\pi\right)  \equiv\frac{\Gamma\left(  \bar{K}%
\rightarrow\pi\pi\right)  -\Gamma\left(  K\rightarrow\pi\pi\right)  }%
{\Gamma\left(  \bar{K}\rightarrow\pi\pi\right)  +\Gamma\left(  K\rightarrow
\pi\pi\right)  }\;\div\;\operatorname{Im}\left(  \frac{A_{2}}{A_{0}}\right)
\sin\left(  \delta_{2}-\delta_{0}\right)  \label{Eq7}%
\end{equation}
within the Standard Model. The CP-conserving $\delta_{I}$ can be extracted
from $\pi\pi$-scattering data at the K-mass scale. On the other hand, the
CP-violating prefactor is now under control thanks to the recently measured
direct CP-violation parameter (\cite{NA48}, \cite{KTeV})%
\begin{equation}
\varepsilon^{\prime}=\frac{i}{\sqrt{2}}\operatorname{Im}\left(  \frac{A_{2}%
}{A_{0}}\right)  e^{i\left(  \delta_{2}-\delta_{0}\right)  }\;. \label{Eq8}%
\end{equation}
So the CP-asymmetries defined in Eq.(\ref{Eq7}) are calculable today. However,
the hadronic parametrization introduced in Eqs.(\ref{Eq4}) was already
sufficient to predict (long time ago) what we will henceforth call quantized
CP-asymmetries%
\begin{equation}
\frac{A_{CP}^{K}\left(  \pi^{0}\pi^{0}\right)  }{A_{CP}^{K}\left(  \pi^{-}%
\pi^{+}\right)  }=-2 \label{Eq9}%
\end{equation}
in the phenomenological limit ($A_{0}>>A_{2}$) of structured hadronic $K^{0}%
$-decay branching ratios%
\begin{equation}
\frac{Br\left(  \pi^{0}\pi^{0}\right)  }{Br\left(  \pi^{-}\pi^{+}\right)
}=\frac{1}{2}\;. \label{Eq10}%
\end{equation}
As we shall see, this seemly academic exercise turns out to be fruitful in
B-physics where numerous decay modes are now available.

\section{Unquantized CP-asymmetries in $B\rightarrow\pi\pi$}

The isospin decomposition of the hadronic B-decays into two pions is identical
to the one already given in Eqs.(\ref{Eq4}). However, comparison of the
inclusive lifetime ratio \cite{PDG2002}%
\begin{equation}
\frac{\tau\left(  B^{+}\right)  }{\tau\left(  B^{0}\right)  }=1.08\pm0.02
\label{Eq11}%
\end{equation}
with Eq.(\ref{Eq2}) strongly suggests that non-factorizable $\Delta B=1$ quark
diagrams (see Fig.1 with $b$ substituted for $s$) are much less efficient at
the B-mass scale. The present $B^{+}/B^{0}$ pattern displayed by exclusive
$B\rightarrow\pi\pi$ branching ratios (\cite{CLEO}, \cite{Belle},
\cite{BaBar}):%
\begin{align}
Br\left(  B^{+}\rightarrow\pi^{0}\pi^{+}\right)   &  =\left(  5.8\pm
1.0\right)  \times10^{-6}\nonumber\\
Br\left(  B^{0}\rightarrow\pi^{-}\pi^{+}\right)   &  =\left(  4.7\pm
0.5\right)  \times10^{-6}\label{Eq12}\\
Br\left(  B^{0}\rightarrow\pi^{0}\pi^{0}\right)   &  =\left(  2.0\pm
0.7\right)  \times10^{-6}\nonumber
\end{align}
tends to confirm this expectation, though no real structure emerges yet. In
particular, the rather large central value for the $\pi^{0}\pi^{+}$ mode
definitely excludes a new $\Delta I=1/2$ rule. On the other hand, the
surprisingly large central value for the $\pi^{0}\pi^{0}$ mode apparently
requires a sizeable CP-conserving phase ($\delta_{2}-\delta_{0}$) \cite{Hou99}
to dilute the fully factorized hierarchy (see Eq.(\ref{Eq5})) predicted by
Fig.1a. More precise measurements might of course change the whole pattern in
$B\rightarrow\pi\pi$. However, at this point we should also seriously question
the validity of the parametrization (\ref{Eq4}) used before arguing for large,
unquantized CP-asymmetries. Let us therefore briefly recall what the
theoretical assumptions behind it are.

Assuming CPT-invariance, the $\mathbf{S}$-matrix for the $P^{0}\rightarrow
\pi\pi$ multiplet of decay amplitudes%
\begin{equation}
W=\left(
\begin{array}
[c]{c}%
P^{0}\rightarrow\pi^{-}\pi^{+}\\
P^{0}\rightarrow\pi^{0}\pi^{0}%
\end{array}
\right)  \label{Eq13}%
\end{equation}
reads%
\begin{equation}
\mathbf{S}=\left(
\begin{array}
[c]{cc}%
1 & iW^{t}\\
iCP(W) & S
\end{array}
\right)  \label{Eq14}%
\end{equation}
with%
\begin{equation}
S=\left(
\begin{array}
[c]{cc}%
\pi^{-}\pi^{+}\rightarrow\pi^{-}\pi^{+} & \pi^{-}\pi^{+}\rightarrow\pi^{0}%
\pi^{0}\\
\pi^{0}\pi^{0}\rightarrow\pi^{-}\pi^{+} & \pi^{0}\pi^{0}\rightarrow\pi^{0}%
\pi^{0}%
\end{array}
\right)  \;. \label{Eq15}%
\end{equation}
Imposing then the unitarity of $\mathbf{S}$, i.e. $\mathbf{S}^{\dagger
}\mathbf{S=SS}^{\dagger}=\mathbf{1}$, for this subset of final states, we
obtain (see \cite{Smith} for more details) the ''generalized Watson theorem''
\cite{Watson}%
\begin{equation}%
\begin{array}
[c]{c}%
W=\sqrt{S}W_{b}\\
CP\left(  W\right)  =\sqrt{S}W_{b}^{\ast}%
\end{array}
\label{Eq16}%
\end{equation}
with $W_{b}$ the bare amplitudes denoted as%
\begin{equation}
W_{b}=\left(
\begin{array}
[c]{c}%
P^{0}\rightarrow\left\{  \pi^{-}\pi^{+}\right\} \\
P^{0}\rightarrow\left\{  \pi^{0}\pi^{0}\right\}
\end{array}
\right)  \;. \label{Eq17}%
\end{equation}
Because bare amplitudes simply get complex conjugated under CP (see
Eqs.(\ref{Eq16})), they obviously do not contain any strong phase. In other
words, the FSI effects are contained in $\sqrt{S}$ and factorize.
Consequently, the bare amplitudes are real (up to CKM factors) and arise from
the quark diagrams projected on specific light hadronic states.

Our restriction to isospin-multiplets is called the SU(2)-elastic hypothesis
since the unitarity of the $\mathbf{S}$-matrix implies then the probability
conservation:%
\begin{equation}
\left|  W\right|  ^{2}=\left|  \sqrt{S}W_{b}\right|  ^{2}=\left|
W_{b}\right|  ^{2}\;. \label{Eq18}%
\end{equation}
The isospin symmetry relates the bare final states to the isospin states%
\begin{align}
\left(
\begin{array}
[c]{c}%
\left|  0,0\right\rangle \\
\left|  2,0\right\rangle
\end{array}
\right)   &  =\underbrace{\left(
\begin{array}
[c]{cc}%
\sqrt{2/3} & \sqrt{1/3}\\
\sqrt{1/3} & -\sqrt{2/3}%
\end{array}
\right)  }\left(
\begin{array}
[c]{l}%
\left\{  \pi^{-}\pi^{+}\right\} \\
\left\{  \pi^{0}\pi^{0}\right\}
\end{array}
\right)  \;.\label{Eq19}\\
&  \;\;\;\;\;\;\;\;\;\;\;\;\;\;\;\;\;\;\;O_{SU\left(  2\right)  }\nonumber
\end{align}
In the isospin state basis, the rescattering matrix $S$ is diagonal%
\begin{equation}
S_{diag}=\left(
\begin{array}
[c]{cc}%
e^{2i\delta_{0}} & 0\\
0 & e^{2i\delta_{2}}%
\end{array}
\right)  \label{Eq20}%
\end{equation}
such that%
\begin{align}
W  &  =\sqrt{S}.O_{SU\left(  2\right)  }^{t}.\left(
\begin{array}
[c]{c}%
A_{0}\\
A_{2}%
\end{array}
\right) \nonumber\\
&  =O_{SU\left(  2\right)  }^{t}.\sqrt{S_{diag}}.\left(
\begin{array}
[c]{c}%
A_{0}\\
A_{2}%
\end{array}
\right)  \label{Eq21}%
\end{align}
and we recover the standard parametrization (\ref{Eq4}).

SU(2)-elasticity follows from imposing a bloc-diagonal form for $S$, each bloc
corresponding to an isospin multiplet. This assumption is obviously reasonable
at the K-mass scale, but certainly questionable at the B-mass scale where so
many channels are open. Systematic cancellations among many final state
rescatterings cannot be excluded and theoretical estimates based on Regge
theory tend to support this picture at the D-mass scale (see
\cite{GerardWeyers}). For $B\rightarrow\pi\pi$ decays, this approach predicts
a rather small phase shift such that large non-factorizable contributions (see
Fig.1b) are needed to enhance the $\pi^{0}\pi^{0}$ mode. Waiting eagerly on
better precision measurements for this mode, we now would like to argue that
the $B\rightarrow\pi K$ decays already provide us with a very interesting
laboratory to test (and also extend) the phenomenological SU(2)-elasticity assumption.

\section{Quantized CP-asymmetries in $B\rightarrow\pi K$}

In $B\rightarrow\pi K$ decays, the well-established CKM mixing hierarchy may
supply for strong dynamics to ensure an effective $\Delta I=0$ rule. Indeed,
charm and top quark contributions in Fig.1b are double-Cabibbo-enhanced%
\begin{equation}
\left(  \sin\theta_{c}\right)  ^{2}\sim\frac{1}{20} \label{Eq22}%
\end{equation}
compared to the other non-factorizable and factorizable diagrams. In that
sense, the $B\rightarrow\pi K$ system is similar to the $K\rightarrow\pi\pi$
one (see Eq.(\ref{Eq3})) and we easily understand the measured branching
ratios (\cite{CLEO}, \cite{Belle}, \cite{BaBar})%
\begin{align}
Br\left(  B^{+}\rightarrow\pi^{0}K^{+}\right)   &  =\left(  12.7\pm1.2\right)
\times10^{-6}\nonumber\\
Br\left(  B^{+}\rightarrow\pi^{+}K^{0}\right)   &  =\left(  18.1\pm1.7\right)
\times10^{-6}\label{Eq23}\\
Br\left(  B^{0}\rightarrow\pi^{0}K^{0}\right)   &  =\left(  10.2\pm1.5\right)
\times10^{-6}\nonumber\\
Br\left(  B^{0}\rightarrow\pi^{-}K^{+}\right)   &  =\left(  18.5\pm1.0\right)
\times10^{-6}\nonumber
\end{align}
which clearly exhibit a structured pattern with $1/2$ factors originating from
the $\pm1/\sqrt{2}$ Clebsch-Gordan coefficients for $\pi^{0}$.

\subsection{SU(2)-elasticity}

Encouraged by such a $\Delta I=0$ rule, let us first proceed as for
$K\rightarrow\pi\pi$ and assume%
\begin{equation}
\pi K\rightleftharpoons\pi K\label{Eq23b}%
\end{equation}
SU(2)-elastic rescatterings. We then have the following hadronic decay
amplitudes:%
\begin{align}
A^{0+} &  =\sqrt{\frac{1}{3}}\left(  A_{1/2}^{\prime}+A_{1/2}\right)
e^{i\delta_{1/2}}+\sqrt{\frac{2}{3}}A_{3/2}e^{i\delta_{3/2}}\nonumber\\
A^{+0} &  =\sqrt{\frac{2}{3}}\left(  A_{1/2}^{\prime}+A_{1/2}\right)
e^{i\delta_{1/2}}-\sqrt{\frac{1}{3}}A_{3/2}e^{i\delta_{3/2}}\label{Eq24}\\
A^{00} &  =-\sqrt{\frac{1}{3}}\left(  A_{1/2}^{\prime}-A_{1/2}\right)
e^{i\delta_{1/2}}+\sqrt{\frac{2}{3}}A_{3/2}e^{i\delta_{3/2}}\nonumber\\
A^{-+} &  =\sqrt{\frac{2}{3}}\left(  A_{1/2}^{\prime}-A_{1/2}\right)
e^{i\delta_{1/2}}+\sqrt{\frac{1}{3}}A_{3/2}e^{i\delta_{3/2}}\nonumber
\end{align}
with $A_{1/2}^{\prime}$ ($A_{1/2},A_{3/2}$) the reduced matrix elements of the
isosinglet (isotriplet) weak Hamiltonian, respectively \cite{GerardWeyersSU2}.

Straightforward projections on appropriate color-singlet states link then the
$A_{1/2}^{\prime}$ bare amplitude to the Cabibbo-enhanced non-factorizable
quark diagrams (see Fig.2a).%
\begin{align*}
&
\begin{array}
[c]{cc}%
%TCIMACRO{\FRAME{itbpF}{1.631in}{0.9098in}{0.3987in}{}{}{fig2a.eps}%
%width 1.631in;  height 0.9098in;  depth 0.3987in;  original-width 2.9153in;
%original-height 1.6042in;  cropleft "0";  croptop "1";  cropright "1";
%cropbottom "0";  filename 'Fig2a.eps';file-properties "XNPEU";}}}%
%BeginExpansion
\raisebox{-0.3987in}{\includegraphics[
height=0.9098in,
width=1.631in
]%
{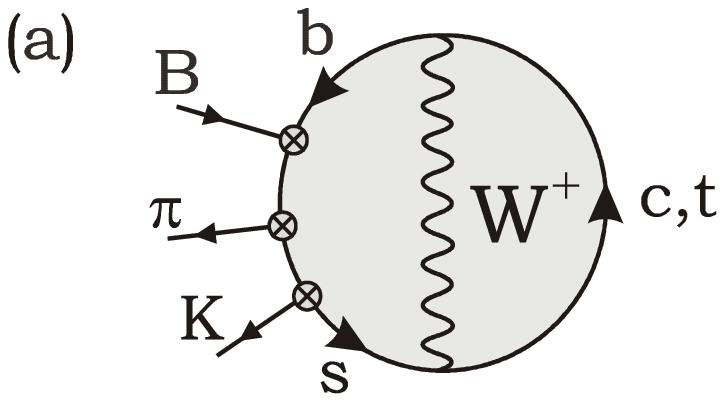}%
}%
%EndExpansion
& \;%
%TCIMACRO{\FRAME{itbpF}{2.5771in}{0.8916in}{0.3831in}{}{}{fig2b.eps}%
%width 2.5771in;  height 0.8916in;  depth 0.3831in;  original-width 4.6363in;
%original-height 1.5714in;  cropleft "0";  croptop "1";  cropright "1";
%cropbottom "0";  filename 'Fig2b.eps';file-properties "XNPEU";}}}%
%BeginExpansion
\raisebox{-0.3831in}{\includegraphics[
height=0.8916in,
width=2.5771in
]%
{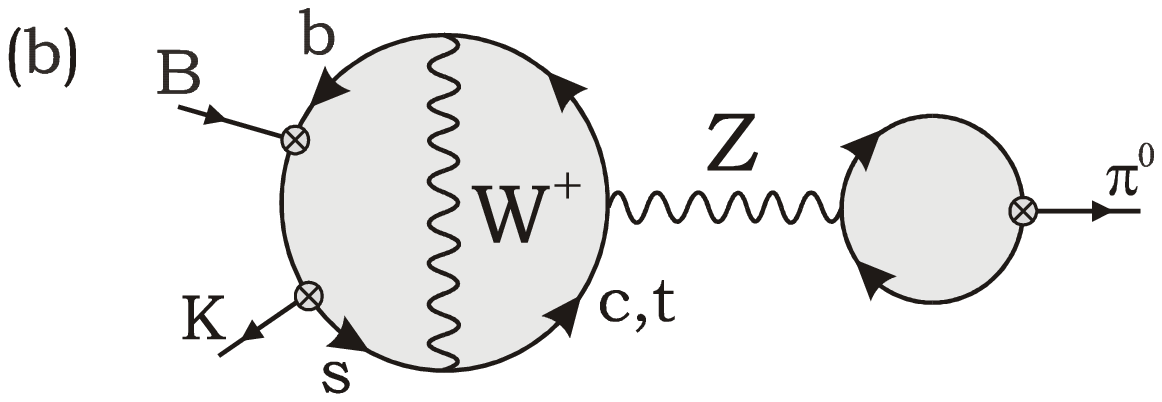}%
}%
%EndExpansion
\smallskip
\end{array}
\\
&  \;\;\;\;\;\underline{Fig.2}\text{\ Double-Cabibbo-enhanced }B\overset
{}{\rightarrow}\left\{  \pi K\right\}  \text{ decay diagrams:}\\
&  \;\;\;\;\;\;\;\;\;\;\;\;\;\;(\text{a})\text{ non-factorizable and
}(\text{b})\text{ hybrid.}%
\end{align*}
Notice that the second order ''hybrid'' diagrams shown in Fig.2b only produce
a neutral pion in the final state. We conclude from Eqs.(\ref{Eq24}) with
$\delta_{I}=0$ that they contribute to the $A_{1/2}$ and $A_{3/2}$ bare
amplitudes, but not to $A_{1/2}^{\prime}$.

In the phenomenological limit ($A_{1/2}^{\prime}>>A_{1/2},A_{3/2}$) of
perfectly structured branching ratios, the direct CP-asymmetries%
\begin{equation}
A_{CP}^{B}\left(  \pi K\right)  \equiv\frac{\Gamma\left(  \bar{B}%
\rightarrow\pi\bar{K}\right)  -\Gamma\left(  B\rightarrow\pi K\right)
}{\Gamma\left(  \bar{B}\rightarrow\pi\bar{K}\right)  +\Gamma\left(
B\rightarrow\pi K\right)  }\;\div\;\operatorname{Im}\left(  \frac{A_{3/2}%
}{A_{1/2}^{\prime}}\right)  \sin\left(  \delta_{3/2}-\delta_{1/2}\right)
\label{Eq25}%
\end{equation}
derived from Eqs.(\ref{Eq24}) are quantized in a way similar to Eq.(\ref{Eq9}%
):%
\begin{equation}
A_{CP}^{B}\left(  \pi^{0}K^{+}\right)  :A_{CP}^{B}\left(  \pi^{+}K^{0}\right)
:A_{CP}^{B}\left(  \pi^{0}K^{0}\right)  :A_{CP}^{B}\left(  \pi^{-}%
K^{+}\right)  =+2:-1:-2:+1\;.\label{Eq26}%
\end{equation}
Such a simple pattern is certainly not excluded by present preliminary data
from CLEO, Belle and BaBar experiments, whose current combined averages are
(\cite{CLEO}, \cite{Belle}, \cite{BaBar})%
\begin{align}
A_{CP}^{B}\left(  B^{+}\rightarrow\pi^{0}K^{+}\right)   &  =-0.10\pm
0.08\nonumber\\
A_{CP}^{B}\left(  B^{+}\rightarrow\pi^{+}K^{0}\right)   &  =+0.05\pm
0.08\label{Eq26b}\\
A_{CP}^{B}\left(  B^{0}\rightarrow\pi^{0}K^{0}\right)   &  =+0.03\pm
0.37\nonumber\\
A_{CP}^{B}\left(  B^{0}\rightarrow\pi^{-}K^{+}\right)   &  =-0.08\pm
0.04\;.\nonumber
\end{align}
The observation of a large departure from the quantized pattern Eq.(\ref{Eq26}%
) would indicate sizeable FSI effects beyond SU(2)-elasticity. Let us
therefore anticipate such a possibility by illustrating how far one can go in
that direction without invoking intricate fits.

\subsection{Beyond SU(2)-elasticity}

As we already said, SU(2)-elasticity requires a bloc-diagonal form for
$\sqrt{S}$ such that states belonging to different iso-multiplets do not
communicate. From a phenomenological point of view, it is interesting to have
a formalism in hand allowing for more general rescatterings. So, let us assume
that B-mesons can also decay into some other unsuppressed $I=1/2$ bare mode
$\left\{  XY\right\}  $ (see Fig.3) which then rescatters into the physical
$\pi K$ states.%
\begin{align*}
&
\begin{array}
[c]{cc}%
%TCIMACRO{\FRAME{itbpF}{1.67in}{0.8951in}{0.4765in}{}{}{fig3a.eps}%
%width 1.67in;  height 0.8951in;  depth 0.4765in;  original-width 2.9853in;
%original-height 1.5766in;  cropleft "0";  croptop "1";  cropright "1";
%cropbottom "0";  filename 'Fig3a.eps';file-properties "XNPEU";}}}%
%BeginExpansion
\raisebox{-0.4765in}{\includegraphics[
height=0.8951in,
width=1.67in
]%
{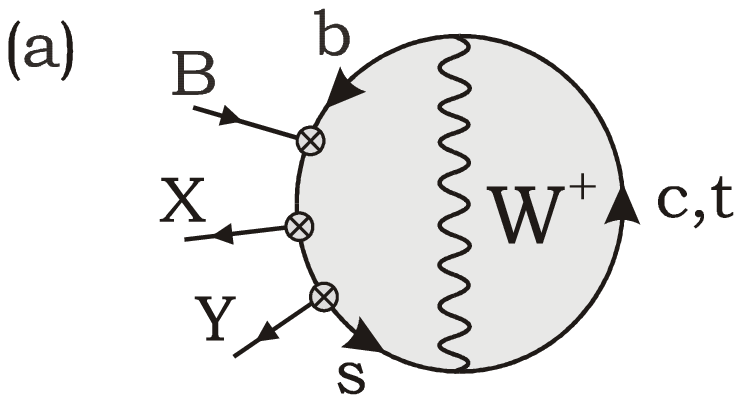}%
}%
%EndExpansion
& \;\;%
%TCIMACRO{\FRAME{itbpF}{2.0885in}{0.7775in}{0.352in}{}{}{fig3b.eps}%
%width 2.0885in;  height 0.7775in;  depth 0.352in;  original-width 3.7464in;
%original-height 1.3638in;  cropleft "0";  croptop "1";  cropright "1";
%cropbottom "0";  filename 'Fig3b.eps';file-properties "XNPEU";}}}%
%BeginExpansion
\raisebox{-0.352in}{\includegraphics[
height=0.7775in,
width=2.0885in
]%
{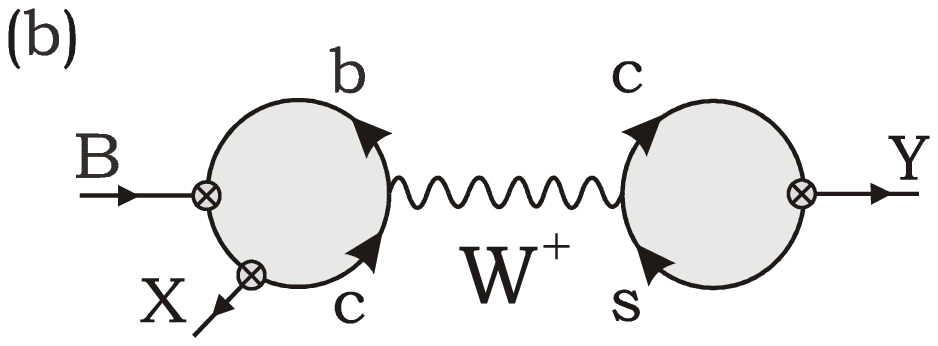}%
}%
%EndExpansion
\end{array}
\\
&  \underline{Fig.3}\text{\ Cabibbo-favored }B\overset{}{\rightarrow}\left\{
XY\right\}  \text{ decay diagrams:}\\
&  \;\;\;\;\;\;\;\;\;(\text{a})\text{ non-factorizable and }(\text{b})\text{
factorizable.}%
\end{align*}
In that case, we have to enlarge the isospin content such that%
\begin{align}
\left(
\begin{array}
[c]{c}%
\left|  1/2^{\prime},\pm1/2\right\rangle \\
\left|  1/2,\pm1/2\right\rangle \\
\left|  3/2,\pm1/2\right\rangle
\end{array}
\right)   &  =\underbrace{\left(
\begin{array}
[c]{ccc}%
1 & 0 & 0\\
0 & \pm\sqrt{1/3} & \sqrt{2/3}\\
0 & \sqrt{2/3} & \mp\sqrt{1/3}%
\end{array}
\right)  }\left(
\begin{array}
[c]{l}%
\left\{  XY\right\}  ^{_{0}^{+}}\\
\{\pi^{0}K^{_{0}^{+}}\}\\
\{\pi^{\pm}K^{_{+}^{_{0}}}\}
\end{array}
\right)  \;.\label{Eq27}\\
&  \;\;\;\;\;\;\;\;\;\;\;\;\;\;\;\;\;\;\;\;\;\;\;O_{SU\left(  2\right)
}^{^{_{0}^{+}}}\nonumber
\end{align}
\underline{Without breaking SU(2)}, we can introduce new rescattering channels
by mixing the two 1/2 representations (see \cite{Smith})%
\begin{align}
\left(
\begin{array}
[c]{c}%
\left|  C_{1}\left(  1/2,\pm1/2\right)  \right\rangle \\
\left|  C_{2}\left(  1/2,\pm1/2\right)  \right\rangle \\
\left|  C_{3}\left(  3/2,\pm1/2\right)  \right\rangle
\end{array}
\right)   &  =\underbrace{\left(
\begin{array}
[c]{ccc}%
\cos\chi & \sin\chi & 0\\
-\sin\chi & \cos\chi & 0\\
0 & 0 & 1
\end{array}
\right)  }\left(
\begin{array}
[c]{c}%
\left|  1/2^{\prime},\pm1/2\right\rangle \\
\left|  1/2,\pm1/2\right\rangle \\
\left|  3/2,\pm1/2\right\rangle
\end{array}
\right)  \;.\label{Eq28}\\
&  \;\;\;\;\;\;\;\;\;\;\;\;\;\;\;\;\;\;\;\;\;O_{\chi}\nonumber
\end{align}
We then obtain%
\begin{align}
W^{_{0}^{+}}  &  =\sqrt{S_{\chi}^{_{0}^{+}}}.O_{SU\left(  2\right)  }%
^{_{0}^{+}\;t}.\left(
\begin{array}
[c]{c}%
B_{1/2}^{\pm}\\
A_{1/2}^{\prime}\pm\,A_{1/2}\\
A_{3/2}%
\end{array}
\right) \nonumber\\
&  =O_{SU\left(  2\right)  }^{_{0}^{+}\;t}.O_{\chi}^{t}.\sqrt{S_{diag}%
}.O_{\chi}.\left(
\begin{array}
[c]{c}%
B_{1/2}^{\pm}\\
A_{1/2}^{\prime}\pm A_{1/2}\\
A_{3/2}%
\end{array}
\right)  \label{Eq29}%
\end{align}
since the rescattering matrix is now diagonal in the $C_{i}$-state basis:%
\begin{equation}
S_{diag}=\left(
\begin{tabular}
[c]{ccc}%
$e^{2i\delta_{C_{1}}}$ & $0$ & $0$\\
$0$ & $e^{2i\delta_{C_{2}}}$ & $0$\\
$0$ & $0$ & $e^{2i\delta_{C_{3}}}$%
\end{tabular}
\ \ \ \ \ \ \right)  \;. \label{Eq30}%
\end{equation}
Let us emphasize that the rescattering remains elastic with respect to the
full set of states $\left(  \left\{  XY\right\}  ,\left\{  \pi K\right\}
\right)  $ since $\sqrt{S_{\chi}}$ is unitary.

If $\chi=0$, we recover the standard SU(2)-elastic parametrization given in
Eqs.(\ref{Eq24}), with the identifications $\delta_{C_{2}}=\delta_{1/2}$ and
$\delta_{C_{3}}=\delta_{3/2}$. On the other hand, if $\chi$ is an arbitrary
angle, we are in general left with two unrelated CP-conserving phase
differences and, consequently, with unquantized CP-asymmetries. However,
$B\rightarrow\pi K$ quantized CP-asymmetries are recovered in two special
limits for the mixing angle $\chi$.

\subsubsection{Large mixing angle}

Let us first consider the case where $\left\{  \pi K\right\}  $ and $\left\{
XY\right\}  $ are generated by hadronization of the same (Cabibbo-enhanced)
non-factorizable quark diagrams (see Fig.2a and Fig.3a, respectively). The
corresponding bare amplitudes $(B_{1/2}^{\pm})^{NF}$ are of the same order as
$(A_{1/2}^{\prime})^{NF}$. In the SU(3) limit (see \cite{Smith}), this amounts
to consider%
\begin{equation}
\left\{
\begin{array}
[c]{c}%
\pi K\rightleftharpoons\pi K\\
\eta_{8}K\rightleftharpoons\pi K
\end{array}
\right.  \label{Eq31b}%
\end{equation}
elastic rescatterings with a large mixing%
\begin{equation}
\tan\chi=-3 \label{Eq32}%
\end{equation}
and only two independent eigenphases:%
\begin{equation}
\sqrt{S_{diag}}=diag\left(  e^{i\delta_{8}},e^{i\delta_{27}},e^{i\delta_{27}%
}\right)  \;. \label{Eq33}%
\end{equation}
The resulting CP-asymmetries proportional to $\sin\left(  \delta_{27}%
-\delta_{8}\right)  $ result from interferences between NF and F bare
amplitudes and obey the following quantization pattern%
\begin{equation}
A_{CP}^{B}\left(  \pi^{0}K^{+}\right)  :A_{CP}^{B}\left(  \pi^{+}K^{0}\right)
:A_{CP}^{B}\left(  \pi^{0}K^{0}\right)  :A_{CP}^{B}\left(  \pi^{-}%
K^{+}\right)  =+2:-\frac{1}{2}:-\frac{3}{2}:+1\;. \label{Eq34}%
\end{equation}
Notice that this SU(3)-elastic pattern is not very different from the
SU(2)-elastic one given in Eq.(\ref{Eq26}). However, we know that the
$\eta_{8}-\eta_{0}$ mixing is not negligible. New disconnected quark diagrams
have to be taken into account and the observed $B\rightarrow\eta^{\prime}K$
branching ratios are fairly large. So this rather academic exercise is
principally a consistency check of the rescattering formalism presented here
to go beyond SU(2)-elastic FSI (see Eq.(\ref{Eq29})).

\subsubsection{Small mixing angle}

Let us turn to the more interesting case where $\left\{  X_{\bar{c}}%
Y_{c}\right\}  $ is generated by hadronization of the (Cabibbo-favored)
factorizable quark diagram (see Fig.3b). Now, the corresponding bare amplitude
$\left(  B_{1/2}\right)  ^{F}$ dominates over $(A_{1/2}^{\prime})^{NF}$ (with
$B_{1/2}\equiv B_{1/2}^{+}=B_{1/2}^{-}$). For illustration, if we consider%
\begin{equation}
\bar{D}D_{s}\rightleftharpoons\pi K \label{Eq35}%
\end{equation}
rescatterings, the $B\rightarrow\bar{D}D_{s}$ branching ratios are indeed of
the order of 1\% \cite{OtherDD}. On the other hand, we expect the mixing angle
$\chi$ to be rather small since $\left\{  \bar{D}D_{s}\right\}  \rightarrow\pi
K$ can only proceed through the annihilation of the $c\bar{c}$ pair into a
light $q\bar{q}$ pair in a semi-inclusive picture. Therefore, the three
independent eigenphases should approximately be those of the isospin basis,
i.e.%
\begin{equation}
\sqrt{S_{diag}}\approx diag\left(  e^{i\delta_{1/2}^{D}},e^{i\delta_{1/2}%
},e^{i\delta_{3/2}}\right)  \;. \label{Eq36}%
\end{equation}

Notice that the SU(4) symmetry is \underline{not} invoked in any sense, seeing
that $\chi$ is now assumed very small (compare with Eq.(\ref{Eq32})). We have
in fact what we may call an \underline{enlarged} SU(2)-elasticity since
charmed mesons decouple and CP-asymmetries obey the SU(2)-elastic quantization
pattern (\ref{Eq26}) in the limit $\chi=0$.

In the complementary limit $\delta_{3/2}-\delta_{1/2}=0$, charmed meson
rescatterings dominate and the CP-asymmetries proportional to $\sin\left(
2\chi\right)  \sin(\delta_{1/2}^{D}-\delta_{1/2})$ result from interferences
between $\left(  B_{1/2}\right)  ^{F}$ and $(A_{1/2}^{\prime},A_{1/2}%
,A_{3/2})^{F}$ bare amplitudes (to leading order in the CKM mixing, the
amplitudes $(B_{1/2})^{F}$ and $(A_{1/2}^{\prime})^{NF}$ have no relative
CP-violating phase). But factorizable quark diagrams cannot produce a neutral
$K^{0}$ in the final state (see Fig.1a). Consequently, the CP-asymmetries obey
a new quantization pattern:%
\begin{equation}
A_{CP}^{B}\left(  \pi^{0}K^{+}\right)  :A_{CP}^{B}\left(  \pi^{+}K^{0}\right)
:A_{CP}^{B}\left(  \pi^{0}K^{0}\right)  :A_{CP}^{B}\left(  \pi^{-}%
K^{+}\right)  =+1:0:0:+1 \label{Eq37}%
\end{equation}
if, again, $B\rightarrow\pi K$ branching ratios are taken perfectly structured.\newline 

The enlarged isospin-invariant formalism based on Eq.(\ref{Eq29}) allows us to
treat pure SU(2)-elastic and $\bar{D}D_{s}$ rescatterings on an equal footing.
A thorough analysis of $B\rightarrow\pi K$ combining these rescattering
processes will be presented elsewhere. For now, let us just emphasize that the
effects of $\bar{D}D_{s}\rightleftharpoons\pi K$ can be factorized and
absorbed into a redefinition of the isospin amplitudes appearing in
Eqs.(\ref{Eq24}). Using $\sin\chi<<1$, Eq.(\ref{Eq29}) is indeed equivalent to
Eqs.(\ref{Eq24}) if the bare amplitude $A_{1/2}^{\prime}$ is replaced by an
effective decay amplitude including now a CP-conserving strong phase:
\begin{equation}
A_{1/2}^{\prime}\rightarrow\left(  A_{1/2}^{\prime}\right)  ^{eff}%
=A_{1/2}^{\prime}-\left(  1-e^{i(\delta_{1/2}^{D}-\delta_{1/2})}\right)  \chi
B_{1/2}\;. \label{Eq39}%
\end{equation}
It is of course tempting to interpret Eq.(\ref{Eq39}) as the hadronic
representation for intermediate on-shell $c\bar{c}$ rescatterings (see
Fig.2a). In this way, the duality correspondence between the quark-level
picture and the hadron-level picture already advocated \cite{GerardHou} for
(semi-) inclusive processes is implemented in the specific case of
$B\rightarrow\pi K$ exclusive decays.

\section{Conclusion}

If the B meson were infinitely heavy, one would argue \cite{MBInfinite} that
there would be no time for the final hadrons to rescatter and all the direct
CP-asymmetries would simply vanish.

For a finite B mass, the $\bar{D}D_{s}\rightleftharpoons\pi K$ elastic scheme
represents an attempt to isolate the dominant (heavy) intermediate
contributions since the measured $B\rightarrow\bar{D}D_{s}$ branching ratios
are of the order of $1\%$.

On the other hand, the simple SU(2) and SU(3)-elastic approximations are more
difficult to justify because of the large energy release in B decays. The $\pi
K$ and $\eta_{8}K$ states are indeed only a small subset of all possible
(light) intermediate states. However, it might be worth recalling that these
approximations do not assert that each specific inelastic channel is small,
but only assume that the \underline{average} over all inelastic channels still
vanishes when the B mass is taken finite. As a matter of fact, the recent
isospin analysis of the measured $B\rightarrow D\pi$ decays \cite{Dpi} does
not contradict this hypothesis.

The present experimental data do not allow us to exclude one of these hadronic
models for final state interactions. Consequently, we propose a
phenomenological test without theoretical prejudice beyond isospin invariance.

In the phenomenologically reasonable limit of perfectly structured
$B\rightarrow\pi K$ branching ratios, we have obtained three possible sets of
quantized CP-asymmetries displayed in Table 1.%
\begin{align*}
&
\begin{tabular}
[c]{|c|l|c|c|c|}\hline
&  & \multicolumn{3}{|c|}{CP-asymmetries}\\\cline{3-5}
& BR & SU(2)-elastic & SU(3)-elastic & On-shell $c\overline{c}$\\\cline{3-4}%
\cline{3-5}%
\multicolumn{1}{|c|}{} &  & $\pi K\rightleftharpoons\pi K$ &
\multicolumn{1}{|c|}{$%
\begin{array}
[c]{c}%
\pi K\rightleftharpoons\pi K\\
\eta_{8}K\rightleftharpoons\pi K
\end{array}
$} & $\bar{D}D_{s}\rightleftharpoons\pi K$\\\hline\hline
\multicolumn{1}{|l|}{$B^{+}\rightarrow\pi^{0}K^{+}$} &
\multicolumn{1}{|c|}{$1/2$} & $+2$ & $+2$ & $+1$\\
\multicolumn{1}{|l|}{$B^{+}\rightarrow\pi^{+}K^{0}$} &
\multicolumn{1}{|c|}{$1$} & $-1$ & $-1/2$ & $0$\\
\multicolumn{1}{|l|}{$B^{0}\rightarrow\pi^{0}K^{0}$} &
\multicolumn{1}{|c|}{$1/2$} & $-2$ & $-3/2$ & $0$\\
\multicolumn{1}{|l|}{$B^{0}\rightarrow\pi^{-}K^{+}$} &
\multicolumn{1}{|c|}{$1$} & $+1$ & $+1$ & $+1$\\\hline
\end{tabular}
\\
&  \underline{Table\;1}\text{\ Structured branching ratios and quantized
CP-asymmetries}\\
&  \;\;\;\;\;\;\;\;\;\;\;\;\;\text{normalized to the }B^{0}\overset
{}{\rightarrow}\pi^{-}K^{+}\text{ mode.}%
\end{align*}

We do not really expect that one of these ''ideal'' quantization patterns for
CP-asymmetries will eventually emerge from future measurements. In particular,
$B\rightarrow\pi K$ branching ratios are not perfectly structured since
Cabibbo-suppressed factorizable quark diagrams (see Fig.1a) and second-order
hybrid quark diagrams (see Fig.2b) do also contribute to them. However, their
relative weight can only be determined through SU(3) arguments which are in
principle beyond the scope of the present work. Yet, extracting the
factorizable contribution from the available $B\rightarrow\pi\pi$ data, we
checked numerically that corrections to Table 1 cannot exceed 20\% in the full
flavor-SU(3) limit, if the CKM angle $\gamma$ is larger or equal to 60$%
%TCIMACRO{\U{b0}}%
%BeginExpansion
{{}^\circ}%
%EndExpansion
$.

To our surprise, relative signs between CP-asymmetries are \underline{not}
sufficient to exclude models for FSI (see Table 1). Nevertheless, we hope that
more precise measurements of CP-asymmetries in $B\rightarrow\pi K$ decays will
allow us to discriminate soon between the various possible scenarii for final
state interaction effects discussed in the present paper. If such is the case,
the CKM angle $\gamma$ will be directly accessible and confidently confronted
with its best-fit value extracted from the standard unitary triangle.

We also hope that our treatment of intermediate charm at the hadronic level
will be of some use for the forthcoming global fits on B-decays into two light
pseudoscalars \cite{RecentAttempts}.\newline 

{\large Acknowledgements}: We wish to thank W.-S. Hou for a stimulating
discussion which led us to investigate ''quantized CP-asymmetries''. Also, we
thank J. Weyers for his useful comments. This work was supported by the
Federal Office for Scientific, Technical and Cultural Affairs through the
Interuniversity Attraction Pole P5/27. C. Smith acknowledges financial support
from the Institut Interuniversitaire des Sciences Nucl\'{e}aires.

\end{document}